\documentclass[reprint,onecolumn]{revtex4-1}
\usepackage{graphicx}
\usepackage{amssymb}
\usepackage{amsmath}
\usepackage{latexsym}
\usepackage{bbold}

\usepackage{bm}

\usepackage{color}

\usepackage{url}
\usepackage{xcolor}
\definecolor{newcolor}{rgb}{.8,.349,.1}

\newcommand{\B} {\bm{B}}

\newcommand{\X} {\bm{X}}
\newcommand{\x} {\bm{x}}

\newcommand{\rhoL} {\bm{\rho}}

\begin{document}

\title{A tight-coupling scheme sharing minimum information across a spatial interface between gyrokinetic turbulence codes}
\author{J. Dominski}
\email{jdominsk@pppl.gov}
\affiliation{Princeton Plasma Physics Laboratory, 100 Stellarator road Princeton, NJ 08543, USA}

\author{S. Ku}
\affiliation{Princeton Plasma Physics Laboratory, 100 Stellarator road Princeton, NJ 08543, USA}
\author{C.-S. Chang}
\affiliation{Princeton Plasma Physics Laboratory, 100 Stellarator road Princeton, NJ 08543, USA}

\author{J. Choi}
\affiliation{Computer Science and Mathematics Division, Oak Ridge National Laboratory, TE USA}
\author{E. Suchyta}
\affiliation{Computer Science and Mathematics Division, Oak Ridge National Laboratory, TE USA}

\author{S. Parker}
\affiliation{Center for Integrated Plasma Studies, Department of Physics, University of Colorado at Boulder, CO 80309, USA}

\author{S. Klasky}
\affiliation{Computer Science and Mathematics Division, Oak Ridge National Laboratory, TE USA }

\author{A. Bhattacharjee}
\affiliation{Princeton Plasma Physics Laboratory, 100 Stellarator road Princeton, NJ 08543, USA }

\begin{abstract}
A new scheme that tightly couples kinetic turbulence codes across a spatial interface is introduced. This scheme evolves from considerations of competing strategies and down-selection. It is found that the use of a composite kinetic distribution function and fields with global boundary conditions as if the coupled code were one, makes the coupling problem tractable. In contrast, coupling the two solutions from each code across the overlap region is found to be more difficult due to numerical dephasing of the turbulent solutions between two solvers.  Another advantage of the new scheme is that the data movement can be limited to the 3D fluid quantities, instead of higher dimensional kinetic information, which is computationally more efficient for large scale simulations on leadership class computers. \end{abstract}

\maketitle

\section{Introduction}

Fusion research will advance to the burning plasma stage with the planned completion and operation of ITER in the next decade. Guided by the understanding obtained from several fusion experiments as well as theory and simulation activities in the world, ITER is expected to attain ten-fold energy gain and realize plasmas that are well beyond the plasma regimes accessible in present and past experiments. It is thus critical for theory and simulation to be reliably predictive so that ITER can achieve its full potential through well-confined and disruption-free operation scenarios. Since fusion plasmas are complex entities, it is not sufficient to understand them piecemeal. The primary focus of this paper is on the development of the first steps that must be taken if we are to realize a Whole Device Model (WDM) of a fusion plasma, based on first principles, that will enable us to understand and predict the performance of existing fusion devices and next step facilities such as ITER.

In 2016, the Department of Energy (DOE) Exascale Computing Program (ECP) launched an ambitious simulation project called ``High-Fidelity Whole Device Modeling of Magnetically Confined Fusion Plasma'', simulating the whole plasma based on first-principles physics from the magnetic axis to the material wall. This is an ambitious challenge, because such a fusion plasma is characterized by multiple spatial and time scales. As a first step, one turns to gyrokinetic theory as a suitable point of departure, except that even in that case, the numerical methods that have been found to be most effective for the core and the edge plasma are different. This is because the underlying physics and plasma properties in these two domains of a turbulent tokamak plasma are quite different. Instead of attempting to describe the entire plasma by a single kinetic code, we have taken the approach of coupling two kinetic codes, one that works optimally in the core plasma, and the other in the edge (or boundary) plasma.

The first phase goal in ECP-WDM has been defined to be a tight kinetic coupling of core and edge micro-turbulence codes. The core and edge turbulence interact and affect each other, leading to possible emergent behavior that cannot be modeled accurately using one or the other code specific. The coupling between the two codes must be tight in order to yield an accurate and self-consistent global solution.  

As a first step, we begin here by reporting the results from tests of the spatial coupling technique obtained by splitting the global gyrokinetic code XGC~\cite{Ku16} into two executables: one for the core region and the other for the edge region.  The strategy of using the same code for two executables located at different spatial regions, and treating them as separate codes, permits us to study the coupling techniques without the possible complication from any differences in the form of the governing equations used, algorithms used, and error from grid interpolation. Communication between the two executables is carried out using the ADIOS software~\cite{ADIOS}.

In the present study, the plasma is modeled with the electrostatic gyrokinetic equations~\cite{Brizard07} in realistic tokamak geometry. In the gyrokinetic equations, particle motion in 6-dimensional (6D) phase space is reduced to 5D motion by analytically averaging out the gyrophase which represents fast gyration in a strong magnetic field. The target problem is further simplified by using the so-called ``delta-f'' scheme, in an effort to identify the most basic issues in the turbulence coupling through a spatial interface. In the ``delta-f'' scheme, the particle distribution function is split into two components $f=f_0+\delta f$, with $f_0$ a Maxwellian background, consistent with an experimental plasma density, temperature and mean flow profiles, and $\delta f$ a perturbation from $f_0$ in the particle distribution function responsible for micro-turbulence. In this study, the $\delta f$ perturbation is solved with the Particle-In-Cell (PIC) technique. Collision and source terms are not included in the present simplified study. The plasma species modeled here are gyrokinetic ions and adiabatic (Boltzmann) electrons. 
 
\section{Coupled equations} 
 In the system of coupled delta-f core and edge codes, a common 
 background $f_0$ is employed in accordance with the specified radial density and temperature profiles. The perturbed particle distribution functions in the core and edge,
 respectively $\delta f^{\rm Core}$ and $\delta f^{\rm Edge}$, 
are combined into a composite one
\begin{equation}
\delta\check{f}\equiv \varpi\,\delta f^{\rm Core}\ +\ (1- \varpi)\,\delta f^{\rm Edge},
\label{eq:f_combined}
\end{equation}
where $\varpi=\varpi(\psi_{gc})$ is a continuous connection function varying between $0$ and $1$ with respect to the gyrocenter radial position $\psi_{gc}$. 
The radial domain is split into three regions:  core ($\varpi=1$), overlap ($0<\varpi<1$), and edge 
($\varpi=0$); see Fig.~\ref{fig:buffer} and the description in the next sub-section. We use a simple linear form of $\varpi$ here. The effects of a higher-order connection function will be studied later.

The electrostatic potential $\check{\phi}$ is obtained by substituting the composite distribution function $\delta\check{f}$, in the global electrostatic Poisson equation for the whole plasma volume as if the coupled system were a single code,
\begin{equation}
\mathcal{L}\check{\phi}=\bar{n}[\delta\check{f}] .
\label{eq:global_poisson}
\end{equation}
The gyrokinetic Poisson operator, $\mathcal{L}$, has coefficients that are dependent on the moments of $f_0$, i.e., plasma density and temperature; see detailed equations in reference~\cite{Ku18,Dominski18}. The source on the right-hand-side is a combination of core and edge charge density contributions that is calculated from the composite $\delta\check{f}$,
\begin{equation}
\bar{n}[\delta\check{f}]=\int d\X dv_\parallel d\mu d\alpha\,\delta[\X+\rhoL-\x] \ 
\Big[\underbrace{\varpi(\X) \delta f^{\rm C}}_{\text{local in core}}+\underbrace{(1-\varpi(\X))\delta f^{\rm E}}_{\text{local in edge}}\Big]\\
=\bar{n}^C [\varpi \delta f] +\bar{n}^E[(1-\varpi)\delta f],
\label{eq:RHS_Poisson}
\end{equation}
where $\X$ is the gyrocenter position vector, $\rhoL$ is the Larmor vector, $\x$ is the particle position vector in the configuration space, $v_\parallel$ is the particle speed scalar parallel to the magnetic field line, 
$\mu$ is the magnetic moment, $\alpha$ is the gyroangle, and $\delta[\cdot\cdot\cdot]$ is the Dirac operator. 
For consistency, it is crucial to have $\varpi$ defined in gyrocenter space, $\varpi=\varpi(\X)$, before applying 
the guiding-center pull-back operation, $\delta[\X+\rhoL-\x]$. 
An important consequence of  Eq.~\eqref{eq:RHS_Poisson} is that the kinetic coupling is enabled by kinetic integration of the code-local distribution functions $\delta f^C$ and $\delta f^E$ in each code, without the necessity for communicating the 5D information between the codes. This has a significant advantage in large-scale computing where the data movement is expensive, because only the 3D field data $\bar{n}^C$ and $\bar{n}^E$ need to be communicated between the codes for solving the field equation consistently with kinetic physics.

The unified, self-consistent field $\check{\phi}$ is then used, in both the core and edge codes, to evolve the distribution function on each side in accordance with the gyrokinetic Vlasov equation
\begin{equation}
\frac{df}{dt}=\frac{\partial f}{\partial t}+\dot{\X}\cdot\frac{\partial f}{\partial \X}+\dot{v}_\parallel\frac{\partial f}{\partial v_\parallel}=0
\ \text{\ \ \ \ \ 
with\ \ \ \ \ 
}\ 
\begin{cases}
\dot\X=\frac{1}{B_{\parallel}^\star}\left(v_\parallel \bm{B}^\star-\mu\nabla B\times \bm{b}-
q\nabla\langle\check{\phi}\rangle_\alpha\times\bm{b}\right)\\
\dot v_\parallel =-\frac{\dot\X}{mv_\parallel}\cdot(\mu\nabla B-q\nabla\langle\check{\phi}\rangle_\alpha)
\end{cases},
\label{eq:GKE}
\end{equation}
Here $q$ the electric charge, 
$m$ the particle mass, $\langle \ \rangle_\alpha$ the gyroaveraging operation, $\bm{b} =\B/B$ with $\B$ the magnetic field and $B$ its magnitude, and $\B^\star$ the effective magnetic field; see  Ref.~\cite{Ku18,Dominski18}. 
In this core-edge coupling scheme, the only boundary condition to be imposed on the coupled system is the particle and solver boundary condition at the outer boundary of the edge code, as if the coupled system is from a single whole-volume code.  For example, the plasma particles are absorbed at the material wall and recycled back into the plasma region as neutral particles, and the electrostatic potential is grounded at the material wall. Importance of enforcing this self-consistency between the unified global field $\check{\phi}$ and the particle distribution function in obtaining the correct turbulence solution across the interface region cannot be overstated.  As is to be shown later, coupling turbulence solutions through an interface using solutions from each code turns out to be problematic.

In this scheme, the time-stepping of the global core and edge distribution functions is achieved by pushing the composite distribution function independently in each code, but using the common global potential field solution. 
Injecting  $\delta\check{f}$ as defined in Eq.~\eqref{eq:f_combined} into the gyrokinetic total derivative operation leads to
\begin{equation}
\frac{d\delta\check{f}}{dt}=\frac{d\varpi\delta{f}^{\rm C}}{dt}+\frac{d(1-\varpi)\delta{f}^{\rm E}}{dt}
=\varpi\frac{d\delta{f}^{\rm C}}{dt}+(1-\varpi)\frac{d\delta{f}^{\rm E}}{dt}
+(\delta f^{\rm C}-\delta{f}^{\rm E})\,\dot{\X}\cdot\nabla\varpi.
\label{eq:weight_evolution}
\end{equation}
If $\delta f^{\rm C}$ and $\delta f^{\rm E}$ are the same at $t=0$ and are evolved using the same gyrokinetic equation, including the coupled self-consistent 
field $\check{\phi}$, then $\delta f^{\rm C}\simeq\delta f^{\rm E}$ locally in the interface region at all times $t>0$, and 
\[
\frac{d\delta\check{f}}{dt}\simeq\varpi\frac{d\delta{f}^{\rm C}}{dt}+(1-\varpi)\frac{d\delta{f}^{\rm E}}{dt}.
\] 
Numerical errors may not guarantee that $\delta f^{\rm C}=\delta f^{\rm E}$ is exactly satisfied, in particular for PIC codes where 
there might be a systematic sampling difference between $\delta f^{\rm C}$ and $\delta f^{\rm E}$. Enlarging 
the overlap region decreases $\nabla\varpi$ and increasing the number of markers decreases the sampling difference on 
$\delta f^{\rm C}-\delta f^{\rm E}$.  This error source in the coupling scheme needs to be carefully examined through convergence study. 

The tight coupling scheme based on these equations is applied at each stage 
of XGC's Runge-Kutta time integrator with a four-step algorithm: 
\begin{enumerate}
\item Prior to solving for the field, both the edge and core simulations compute their contribution to the charge density, Eq.~\eqref{eq:RHS_Poisson}. 
\item The core simulation sends its charge density contribution to the edge simulation. 
\item The edge solves the gyrokinetic Poisson equation globally including the core region, 
Eq.~\eqref{eq:global_poisson}, and sends the field $\check{\phi}$ to the core code.  We use the edge solver here because, the edge code is usually more general and can cover the whole volume.  The field solver could be a separate routine. Using only one global field solver guarantees that the exact same field is 
used by both codes.
\item Both simulations push their $\delta f$ with the same gyrokinetic Vlasov equation, by using exactly the same global field $\check{\phi}$ distribution, Eq.~\eqref{eq:GKE}. 
\end{enumerate}

\section{Radial domain decomposition with overlap and buffer regions}
\begin{figure}
\begin{center}
\includegraphics[width=9cm]{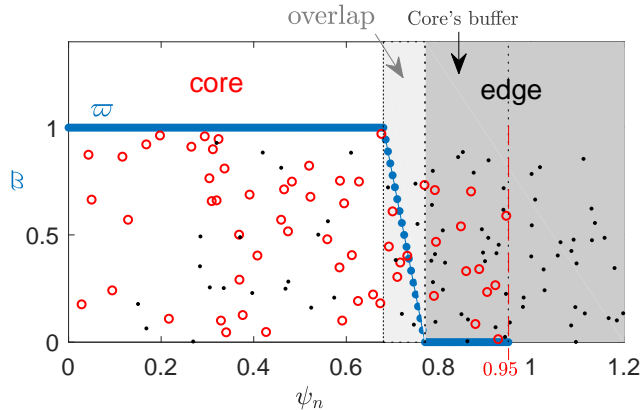}
\end{center}
\caption{Illustration of the radial domains of coupled simulation. The core, the overlap and the edge regions are shown.  The connection function $\varpi$ is described in the three regions: in the core $\varpi=1$, 
in the {\it overlap region}  $0<\varpi<1$, and in edge $\varpi=0$. The core simulation domain is going beyond the overlap up to 
$\psi=0.95$ (red dashed line). This region of the core simulation where $\varpi=0$ is a buffer region where core simulation's 
markers (red circles) do not contribute to Poisson equation but are nonetheless evolved with their weight according to the gyrokinetic equation. 
A similar buffer region exists in the edge simulation.}
\label{fig:buffer}
\end{figure}
The interface between the core and edge codes is an {\it overlap} region where $0<\varpi<1$. As detailed in the previous section,  
both codes consistently contribute in this overlap region to the Poisson equation's source term defined by Eq.~\eqref{eq:RHS_Poisson} and consistently respond to the Poisson-solved field through Eq.~\eqref{eq:GKE}.

In addition to the overlap region, each of the core and the edge simulation modules have {\it buffer} zones. 
The buffer zones for a particular module is spatially contiguous with the overlap region, and extends into the spatial region for the other module.  The buffer region for a particular module accommodates the particles that travel outside of the interface region into the other region, and return. The existence of these buffer regions also ensure continuity of the turbulence fluxes at the inner and outer edges of the overlap region.  In these regions, marker particles are evolved 
with the coupled Eq.~\eqref{eq:GKE} but do not contribute to the Poisson equation source term, Eq.~\eqref{eq:RHS_Poisson}. 
In the case shown in Fig.~\ref{fig:buffer}, the buffer region of the core module is located between the outer edge of the interface region ($\psi=0.76$) and $0.95$, where $\varpi=0$. In the present study, we have chosen the buffer region of the edge simulation to extend all the way to the magnetic axis from the inner edge of the overlap region, but it can be chosen to have a more limited spatial extent.  Since the plasma volume decreases with $r^2$ in a torus, with $r$ being the minor radius, extension of the inner buffer region all the way to the magnetic axis does not increase significantly the computational cost but increases the accuracy of the coupling.  For the purpose of the present study, including only the outer buffer region will be sufficient for the study of the effect of this region on code coupling. 

The outer buffer region for the core code cannot be extended all the way to the material wall because the equations underlying the core code become invalid near the magnetic separatrix surface and in the scrape-off layer. Thus, a finite radial width is needed, separating the outer buffer region from the magnetic separatrix.  If a particle from the core region exits the outer buffer region or equivalently, exits the simulation region of the core code into the edge region, it is re-injected back to the buffer region along the gyrocenter orbital path in a way which preserves density, momentum, and energy for each marker particle. 

\section{Coupled simulation with a field solver}
\label{sec:coupledSimulation}
\begin{figure}
\begin{center}
\includegraphics[width=14cm]{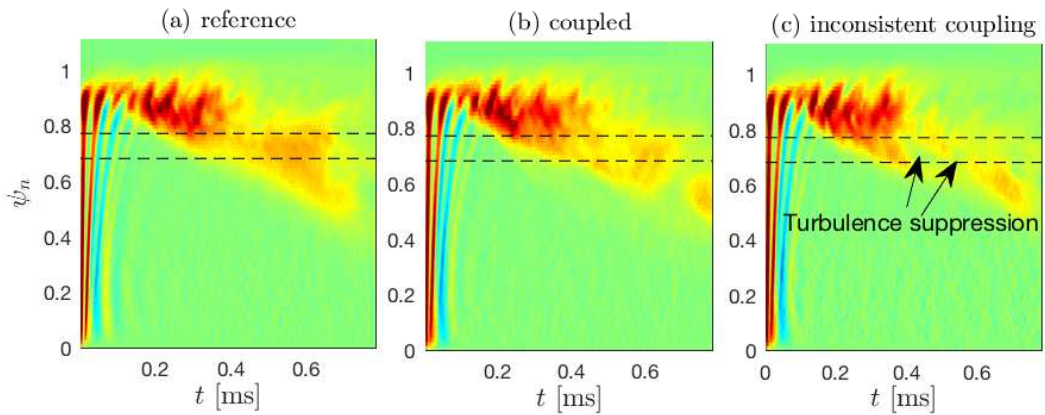}
\end{center}
\caption{Ion heat flux, $Q_i$, measured with a reference XGC simulation through out the whole core-edge volume (a), a coupled one (b), and an example of failed coupling (c). 
These plots show $Q_i$ with respect to the time and radial position $Q_i(t,\psi)$. Horizontal dashed lines 
indicate the position of overlap. The overlap is also indicated in the reference simulation for information. }
\label{fig:flux_comparison_light}
\end{figure}

\begin{figure}
\begin{center}
\includegraphics[width=11cm]{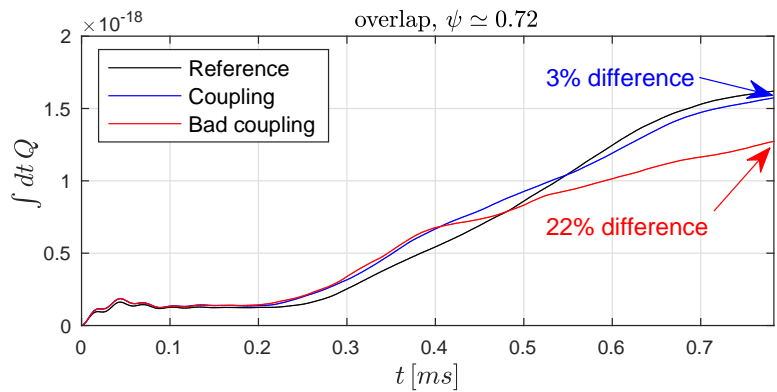}
\end{center}
\caption{Time trace of time-accumulated $Q_i$ measured at $\psi_n=0.72$ at the overlap region. The reference (black line) and coupled simulation (blue line) are in good agreement with a difference smaller than experimentally measurable error bar (normally greater than 10-20\%).}
\label{fig:flux_comparison_light2}
\end{figure}

We demonstrate the fidelity of our coupling scheme, 
by comparing in Fig.~\ref{fig:flux_comparison_light} the results obtained with a reference simulation (a), a coupled simulation (b) and the time-integrated heat-flux across the overlap region in Fig.~\ref{fig:flux_comparison_light2}. Specifically, in these figures we choose to use the ion heat flux, 
$$Q_i(\psi;t)=\left\langle\int dv \,(v_\parallel^2/2+\mu B)f_i \, \dot{\X}\cdot\nabla\psi/|\nabla\psi|\right\rangle_S,$$ with $\langle\ \rangle_S$ defines surface averaging. Heat flux
 is one of the main quantities of great interest in measuring the performance of a magnetically confined fusion plasma. The geometry of our equilibrium is a DIII-D-like plasma including an X-point. 
The ion temperature profile is chosen in such a way that it produces turbulence in the edge region, with a strong temperature gradient just outside of the overlap region. It is well-known that turbulence then propagates radially inward \cite{Chang09}.  There is no heat source in the simulation.  Thus, the temperature gradient relaxes as a result of the turbulent transport, and the turbulence eventually decays. For the boundary condition of the core-edge coupled simulation, the electrostatic potential is zero and particles are absorbed at the material wall outside the magnetic separatrix surface.  For the present simulation, the material wall is placed along a surface that conforms to a magnetic surface at $\psi_n=1.11$. 

In Fig.~\ref{fig:flux_comparison_light} subplot (b), it is shown that the turbulence in the coupled simulation can propagate freely from the edge to the core domain, \textit{i.e.}, from one 
executable to the other, with a high level of fidelity for the problem. To be more quantitative, as measured in Fig.~\ref{fig:flux_comparison_light2} at the end of the simulation, there is only a $3\%$ difference in the time-accumulated turbulent heat flux, $\int dt Q_i(\psi;t)$, passing through a surface in the overlap region, which is the most sensitive region for the coupling challenge. For perspective, it is worth keeping in mind that experimental error bars, against which we eventually need to validate the simulation results, are usually greater than 10-20\%. 

We remark on a notable result of the present study that may appear counter-intuitive. It may appear upon first glance that the most straightforward mathematical way to couple two codes is by having each code use its own solvers and outer boundary conditions, with the solutions matched at the inner interface. Our finding is that this is not so, and that the spatial coupling becomes much easier if two codes solve for the same composite solution, subject to the boundary conditions as if the two codes were one.  As an example, we present here a case which uses separate solvers, with the boundary condition for the core code at the outer end of the buffer region $\psi_n=0.95.$  Since $\psi_n=0.95$ is located multiple radial-orbit-widths and turbulence-correlation-lengths away from the core region $\psi_n < 0.76$, the imposition of a separate solver boundary condition for the core code does not affect significantly the coupled solution.  The result of this case is shown in the subplots (c) 
of Fig.~\ref{fig:flux_comparison_light}, and the red line in Fig.~\ref{fig:flux_comparison_light2}, which is from the following two boundary conditions for the solvers executed in each codes: 
\begin{equation}
\begin{cases}
\mathcal{L}^{\rm C}\phi^{\rm C}=\bar{n}[\varpi f^{\rm C} + (1-\varpi)f^{\rm E}],\  \psi_n\in[0,0.95]\ \text{and}\ \phi^{\rm C}|_{\psi_n=0.95}=0,\\
\mathcal{L}^{\rm E}\phi^{\rm E}=\bar{n}[\varpi f^{\rm C} + (1-\varpi)f^{\rm E}],\  \psi_n\in[0,wall]\ \ \text{and}\ \phi^{\rm E}|_{wall}=0.
\notag
\end{cases}
\end{equation}

As can be seen from Fig.~\ref{fig:flux_comparison_light} (c), turbulence is partially suppressed in the overlap region, and the heat flux through it is much lower than the reference case (Fig.~\ref{fig:flux_comparison_light2}). A detailed study reveals that this is caused by the phase mismatch between the core and edge solutions of the fluctuating turbulent potential field $\phi^C$ and $\phi^E$, resulting in a partial cancellation between them.  An artificial solver boundary condition at $\psi_n=0.95$ is causing this phase mismatch, by not allowing for the phase information to be correctly communicated between the core and edge regions.



\section{Investigation on the influence of the overlap width in a simplified geometry}
\begin{figure}
\begin{center}
\includegraphics[width=15cm]{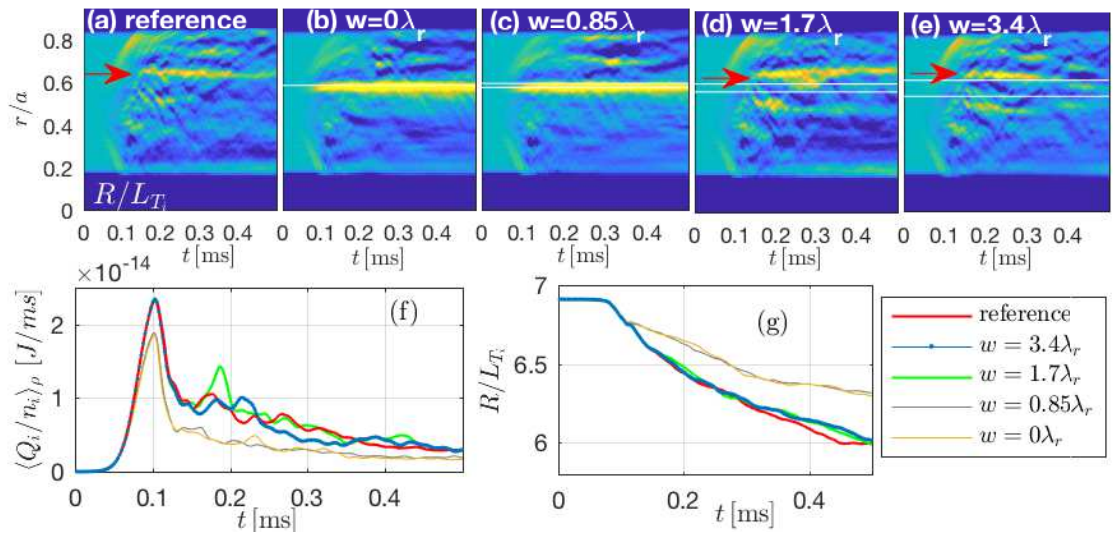}
\end{center}
\caption{Influence of the width of the overlap region in nonlinear coupled simulations: 
Time evolution of the radial profile of temperature gradient 
$R/L_{T_i}$ is plotted in subplots (a,b,c,d,e) for various overlap width.  Time traces of the radially averaged heat flux $Q_i/n$ and the temperature gradient  $R/L_{T_i}$ are plotted in (f) and (g), respectively. The radial average is carried out from $r/a=0.2$ to $0.8$. White horizontal lines indicate the overlap region and $w$ its width in units of radial correlation length of turbulence $\lambda_r$, where we estimated $\lambda_r\simeq2.3$ cm. The radial direction is the radius, $r$, normalized to the radius at boundary $a\simeq59$ cm, such that $r/a=1$ at the edge. A total of $320$M ion marker particles is used in each simulation. }
\label{fig:SCAN_in_Lw_320M}
\end{figure}

Another significant finding from our study is that the overlap width, $w$, must be larger not only than the radial orbital width but also the radial turbulence correlation length, $\lambda_{r}$. 
To study this problem, we carry out a nonlinear simulation scan with respect to the overlap width, from $w=0\times\lambda_r$ to $w\simeq3.4\times\lambda_r$. $\lambda_{r}$ is evaluated to be $2.3$ cm in the overlap region from the reference simulation.
To avoid possible complications from the flux-surface shape, we use a circular cross-section torus, called ``cyclone-base-case geometry'' with the temperature gradient $R/L_{T_i}=6.9$ and the density gradient $R/L_N=2.2$. 
No heat source is applied, such that the plasma is left free to relax through the growth of $\delta f$ via the turbulent heat flux down the gradient. 
For this relaxation problem, we measure the ion temperature gradient $R/L_{T_i}$, plotted in the first row of Fig.~\ref{fig:SCAN_in_Lw_320M}, as well as the radial average of the ion heat flux and the ion temperature gradient in the second row. 
The ion temperature gradient relaxation is the result of the ion heat flux integrated over time and is a very good indicator of the accuracy of the coupling.  Towards the end of the simulation, since the temperature gradient has relaxed and the turbulence is weak, the comparison is less meaningful. 

The reference and the coupled simulations agree well when the overlap region width, $w$, is wide enough. 
In our scan, the best agreement is obtained 
when $w$ is a few times the local correlation length of turbulence, $\lambda_r$. Even when $w\simeq 1.7\lambda_r$, there is a good agreement between the coupled (green) and the reference (red) cases in the radial-averaged heat flux and temperature gradient in subplots (f) and (g), respectively,. However, the zonal structure on $R/L_{T_i}$ near $r/a\simeq0.7$ (red arrows) of the reference simulation (a) shows better agreement with the results 
with (e) $w\simeq 3.4\lambda_r$ than with (d) $w\simeq 1.7\lambda_r$.

\section{Conclusion}
A new tight-coupling scheme is introduced that enables spatial coupling of large-scale kinetic turbulence codes across an interface. Several key points of this scheme are explained. Distribution functions from the core and edge codes are combined into one by using a connection function at the overlap region. Using this connection function, the charge density is calculated from the kinetic information obtained from each code and communicated to the solver routine, whereby the two 3D charge density distributions are combined into one to determine the source term on the right hand side of the global Poisson equation. This method allows us to avoid expensive communication of large-scale 5-dimensional kinetic data.  For successful coupling, the key is that the global Poisson equation and the boundary condition needed to solve it are as if the coupled codes are one. Furthermore, it is found that the overlap region at the coupling interface needs to be of similar size or wider than the radial correlation length of turbulence and radial orbit width. 

In related work, using simplified model equations, Ricketson et al.~\cite{Lee17} have recently shown that classical additive Schwarz-type methods hold considerable promise for addressing coupling problems of the kind considered in this paper. They obtain analytical as well as numerical results on convergence for their model problem. It would be worthwhile to consider using similar techniques why the method proposed in this paper is effective.


Spatial code coupling is an active research area. As the next step in the ECP WDM project, the present scheme is being successfully extended to coupling two different core-edge gyrokinetic codes (GENE ~\cite{Goerler11} in the core and XGC in the edge), which involves grid mapping as an extra ingredient.  This work will be reported in the near future.

\section*{Supplementary material}
See supplementary material for a video comparing the coupled and reference nonlinear simulations presented in Sec.~\ref{sec:coupledSimulation}. The top left and top right subplots represent the electrostatic potential $\phi$ of the coupled and reference simulations, respectively. The overlap region is identified between white lines in top left subplot. The bottom subplot represents the accumulated heat flux of the reference (blue) and coupled (red) simulations through a surface inside the overlap region, $\psi_n\simeq0.72$.

\section*{Acknowledgments}
{
This research was supported by the Exascale Computing Project (17-SC-20-SC), a collaborative effort of the U.S. Department of Energy Office of Science and the National Nuclear Security Administration. 

This research used resources of the Oak Ridge Leadership Computing Facility at the Oak Ridge National Laboratory (ORNL) and resources of the National Energy Research Scientific Computing Center (NERSC), which are supported by the Office of Science of the U.S. Department of Energy under Contracts No. DE-AC05-00OR22725 and DE-AC02-05CH11231, respectively.

This manuscript is based upon work supported by the U.S. Department of Energy, Office of Science, Office of Fusion Energy Sciences, and has been authored by Princeton University under Contract Number DE-AC02-09CH11466 with the U.S. Department of Energy.
}


\bibliography{bibliography}

\end{document}